\documentstyle[prl,aps,twocolumn,epsf,rotate]{revtex}
\newcommand{\be}{\begin{eqnarray}}
\newcommand{\ee}{\end{eqnarray}}

\begin{document}
\title{Old and New Facets of y-Scaling: the Universal Features of Nuclear Structure
Functions and Nucleon Momentum Distributions} 
\author{Claudio Ciofi degli Atti}
\address{
Dipartimento di Fisica,
Universit\`a di Perugia and Istituto Nazionale di Fisica Nucleare,
Sezione di Perugia\\
Via A.Pascoli, I-06100 Perugia, Italy \\
}
\author{Geoffrey B. West}
\address{Theoretical Division, T-8, MS B285, Los Alamos National Laboratory, Los Alamos, NM 87545, USA}

\maketitle

\begin{abstract}
Some systematic general features of y-scaling structure functions, which are essentially
independent of detailed dynamics, are pointed out. Their physical interpretation in terms
of general characteristics, such as a mean field description and nucleon-nucleon
correlations, is given and their relationship to the momentum distributions 
illustrated. A new scaling variable is proposed which allows a direct expression of the
scaling function in term of an integral over the momemtum distributions thereby avoiding
explicit consideration of binding corrections. 
\end{abstract}

\narrowtext
\vskip 1cm

Inclusive quasi-elastic electron scattering is a powerful method
for  measuring the momentum distribution of nucleons inside a nucleus. This is
most  succinctly made manifest by expressing the data in terms of the scaling
variable $y$ which, over a large kinematic range, can be identified as the
longitudinal momentum  of the struck nucleon, $k_{\parallel}$ \cite{west}. 
At sufficiently
large momentum transfers,  $\bf{q}$, the structure
function, $W(\nu, q^{2})$, which represents the deviation of the cross-section
from scattering from free nucleons, scales to a function of the single 
variable $y$ according to $qW(\nu , q^{2}) \approx f(y)$ where $\nu$ is the 
electron energy loss 
and $q\equiv \vert\bf{q}\vert$. Thus, in the scaling limit, $qW$ approaches a 
function that effectively traces out the longitudinal momentum distribution of 
the nucleons:
\begin{equation}
f(y) = \int n(k_{\parallel},{\bf{k}_{\perp}}) d^2 {\bf{k_{\perp}}}
     =   2\pi\int\displaylimits_{\vert y\vert}^{\infty} n(k) k dk
\label{one}
\end{equation}  
Here, $n(k)$ is the conventional nucleon momentum distribution function  
normalized such that
\begin{equation}
\label{three}
\int{d^{3}k}n(k) = \int^{\infty}_{-\infty}dy f(y) = 1
\end{equation}
Knowledge of $f(y)$ can therefore be used to obtain $n(k)$ by inverting Eq.~(\ref{one}):
\begin{equation}
\label{four}
n(k) = - \frac{1}{2 \pi y} \frac{d f(y)} {dy} \hspace{1in}   \vert y\vert = k
\end{equation}
More generally, $qW(\nu, q^{2})$ is related to the spectral function,
$P(k,E)$, which depends on the energy ($E$) as well as the momentum
of the nucleons \cite{ciofi}: 
$qW(\nu , q^{2}) \approx F(y) = f(y) - B(y)$ where 
\begin{equation}
\label{six}
B(y) = 2 \pi \int\displaylimits_{E_{min}}^{\infty}dE   
            \int\displaylimits _{\vert y\vert}^{k_{min}(y,E)}  {P_{1} (k,E)}
\end{equation}
$P_{1}$ is that part of $P(k,E)$ generated by 
ground state correlations; (thus, in a mean field description or, for the case of $^{2}H$,
$P_{1}=0$)\cite{foot}. 

Over the past several years there have been vigorous efforts to explore $y$-scaling over a wide range of
nuclei \cite{day}. The purpose of this paper is to point out, and give some
insight into, some general universal features of $f(y)$ that are essentially
independent of the detailed dynamics and  which seem to have been overlooked in past
discussions. We shall show that the  overall structure and systematics of the data
can, to a large extent, be understood in terms of some rather general
characteristics of nuclei. We provide a physical interpretation of these features
and show their  relationship to momentum distributions. In addition, we shall
illustrate a  new approach which allows a determination 
of $f(y)$ directly from experimental data, thereby avoiding use of 
the theoretical ``binding correction", $B(y)$. Up to now, longitudinal momentum distibutions have been obtained by extracting $F(y)$ from data and then estimating $B(y)$ theoretically. Based on such a procedure a systematic analysis, to be presented elsewhere \cite{faralli}, exhibits the following 
general features of $f(y)$ for nuclei with $A < 56$:

\begin{itemize}

\item[i)] $f(0)$ decreases monotically with $A$, from  $\sim 10 MeV^{-1}$
when $A=2$ to $\sim 3 MeV^{-1}$ for heavy nuclei; moreover, for $y \sim 0$, $f(y) \sim (\alpha^{2} + y^{2})^{-1}$, with $\alpha$ 
ranging from $\sim 45 MeV$ for $A=2$, to $\sim 140 MeV$ for $A=56$.

\item[ii)] For $50MeV \le  \vert y\vert \le 200 MeV$,  $F(y) \sim e^{-a^{2}y^{2}}$ with
$a$  ranging from $\sim 50 MeV$ for $A=2$, to $\sim 150 MeV$ for $A=56$.

\item[iii)] For $\vert y\vert \ge 400 MeV$,  $f(y) \sim B e^{-b \vert y \vert}$, with $B$ 
ranging from $2.5 \times 10^{-4} MeV ^{-1}$ for $A=2$, 
to $2 \times 10^{-4} MeV ^{-1}$ for $A=56$, 
and, most intriguingly, $b = 6 \times 10^{-3} MeV ^{-1}$,  \it{independent} of $A$.
\end{itemize}

The following simple
form for $f(y)$ yields an excellent representation of these general
features: \begin{equation}
\label{seven}
f(y) = \frac{Ae^{-a^{2}y^{2}}}{ \alpha^{2} + y^{2}}  + B e^{-b \vert y \vert}
\end{equation}
The first term dominates the small $y$-behavior whereas the second term dominates 
large $y$.  Let us now discuss the motivation and interpretation of this
equation. The systematics of the first term are determined by the small and
intermediate  momentum behaviour of the single particle wave function. For
$\vert y\vert\le\alpha$ this can be straightforwardly  understood in terms of a zero range
type of approximation and is, therefore,  insensitive to the details of the
microscopic dynamics, or of a specific model. The long  distance behavior of a
single particle wave function is controlled by its separation energy  $(Q\equiv M + M_{A-1} - M_A)$  and is given by $e^{-\alpha r}\over r$ where  $\alpha = (2\mu
Q)^{1\over 2}$, $\mu$ being  the reduced mass of the nucleon.  In
momentum space (see Eq.~(\ref{fivea}) below) this translates into $(k^{2}+\alpha^{2})^{-1}$ so the parameter 
$\alpha$ occurring in (\ref{seven}) is to be identified with $(2\mu Q)^{1\over 2}$.
This agrees well with fits to the data summarized in (i) above. 

Before discussing the intermediate range it is instructive to consider first the 
large $y$-behavior. Perhaps the most intriguing phenomenological characteristic of
the data is that  {\it $f(y)$ falls off exponentially at large $y$ with a similar
slope parameter for all  nuclei, including the deuteron}. Unlike
the behavior for small $y$ there is, as far as we are aware, no simple argument to
explain this remarkable fact. Since (i) $b$ is almost the same for all nuclei 
{\it including} $A=2$, i.e.,  $f(y)$, at large $y$, appears to be simply the rescaled 
longitudinal scaling function of the deuteron; and (ii)  
$b (\approx 1.18 fm) \ll 1/\alpha_D (\approx 4.35 fm)$, we conclude that the term
$e^{-b\vert y \vert}$ is related to 
the short range part of the deuteron wave function and reflects the universal nature 
of $NN$ correlations in nuclei. In momentum space it can be related to the effective single-particle potential, $V(k)$, by 
\begin{equation}
(k^2 + \alpha^2)\Psi ({\bf k}) = 2M\int {d^3k'\over(2\pi)^3}V({\bf k}-{\bf k'})        \Psi ({\bf k'})
\label{fivea}
\end{equation}              
where $\Psi ({\bf k})$ is the single-particle wave function; ($n(k)=|\Psi ({\bf k})|^2$). The relationship of the exponential fall-off to $V({\bf k})$ will be discussed in some detail in a later paper. 
Notice, incidentally, that once $B$ and $b$ are
determined by fitting the large $y$ data, the remaining two parameters in Eq.~(\ref {seven}), $A$ and $a$,
can be fixed, respectively, from the value of $f(0)$ and the normalization condition,
Eq.~(\ref{three}). It is important to stress that, once this is done, there are no
adjustable parameters for different nuclei.

The intermediate range is clearly sensitive to the parameter $a$, the gaussian form 
being dictated by the usual harmonic oscillator potential used in the shell model.
Notice, however, that the gaussian is modulated by the correct $\vert y\vert <\alpha$
behaviour, namely $(y^2 + \alpha^2)^{-1}$, thereby ensuring that the wave function has the correct asymptotics. As an example, Fig.~1 shows the experimental scaling functions, $f(y)$, for $^2H$
and $^4He$ compared to Eq.~(\ref{seven}). The fit is excellent; for $^4He$, the value
of $a$ is slightly smaller than the conventional one obtained from a pure harmonic oscillator potential since the rms radius receives an additional contribution from the term $(\alpha^2 + y^2)^{-1}$ \cite{faralli}. A systematic analysis for a large body of nuclei exhibiting the same
features as those shown in Fig.~1 will be presented elsewhere.

%
\begin{figure}
\vspace{.35cm}
\centerline{\rotate[r]{\epsfxsize=0.5\hsize\epsfbox{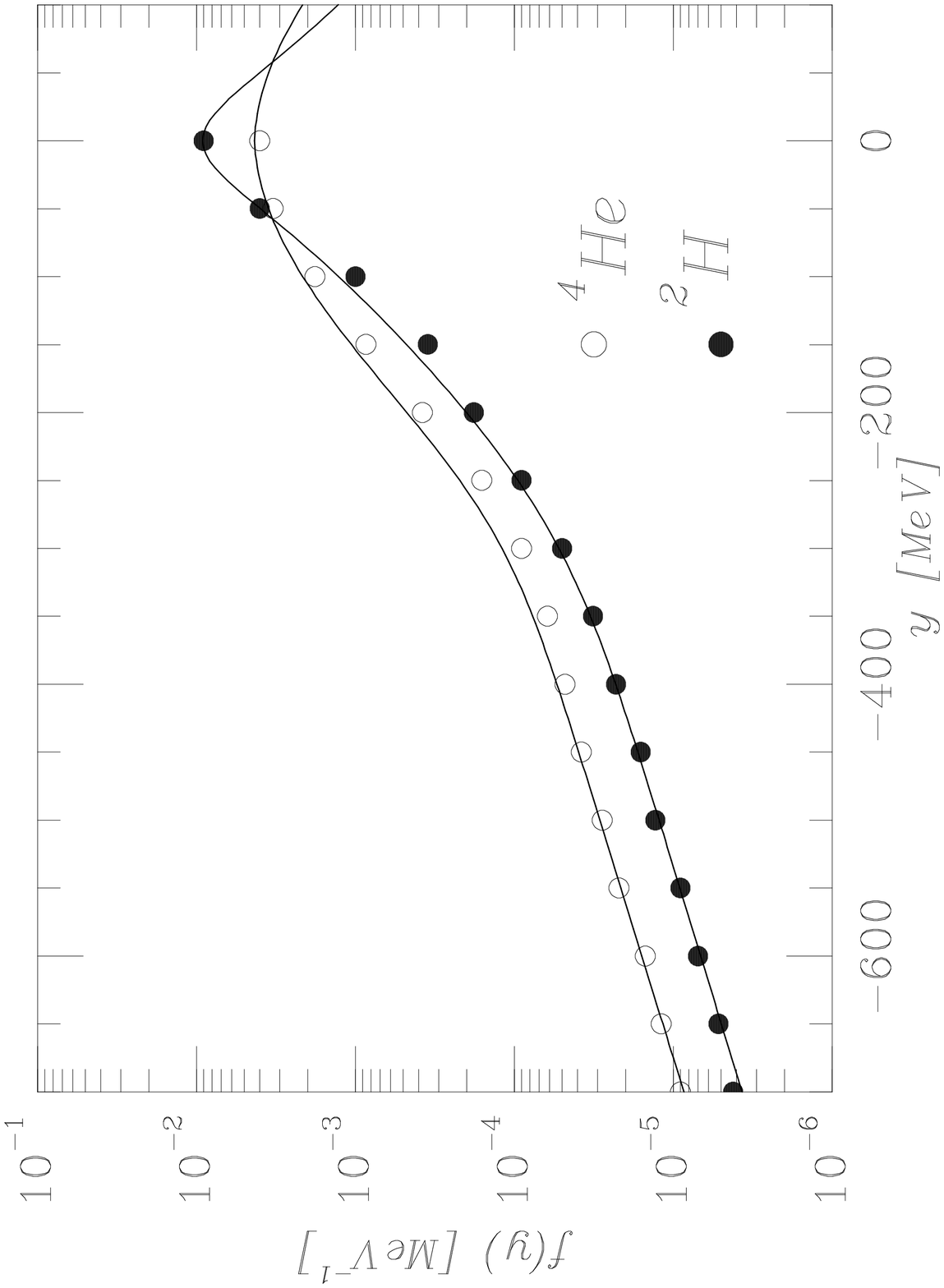}}}
\medskip

{Fig. 1. {\small{The "experimental" longitudinal momentum distributions 
of $^2H$ \cite{ciofi} and $^4He$ \cite{faralli} compared
with Eq.~(\ref{seven}) with $\alpha = 0.23fm^{-1}$ and $B = 0.048$fm for $^2H$ and 
$\alpha = 0.85fm^{-1}$ and $B = 0.12$fm for $^4He$. For both nuclei $b = 1.2fm$. Note that the errors on $f(y)$ for $^2H$ are of the same size as the data points; for $^4He$ they are somewhat larger \cite{faralli}}}}\\

\end{figure}
%


With these observations it is now possible to understand the normalization and 
evolution of $f(y)$ with $A$. First note that Eq.~(\ref{one}) implies
\begin{equation}
\label{nine}
f(0) = {1\over 2}\int {d^3 k} {n(k)\over k}
\end{equation}
In other words, $f(0)$ is a measure of $\langle 1/2k \rangle$ and so, as expected, 
is sensitive to the small momentum, or large distance, behavior of the wave 
function. Now, typical mean momenta vary from around 50 MeV for the deuteron up to 
almost 300 MeV for nuclear matter.  We can, therefore,
immediately see why $f(0)$ varies from around 10 for the deuteron down to around 
2-3 for heavy nuclei \cite{f0}. Since $f(y)$ is
constrained by a sum rule, Eq.~(\ref{three}), whose normalization is independent of 
the nucleus, a decrease in $f(0)$ as one changes the nucleus must be compensated 
for by a spreading of the curve for larger values of $y$.  {\it Thus, an understanding 
of $f(y)$ for small $y$ coupled with an approximately universal fall-off for large $y$, 
together with the constraint of the sum rule,
leads to an almost model-independent understanding of the gross features of the data 
for all nuclei}.

To sum up, the ``experimental" longitudinal momentum distribution can be thought of as
the incoherent sum of a mean field contribution,
$(f_{0} = \frac{A}{\alpha^{2} + y^{2}} e^{-a^{2}y^{2}})$, with the correct
model-independent small $y$-behaviour built in, and a ``universal"
correlation  contribution $(f_{1} = Be^{-b|y|})$. Thus, the momentum distribution,
$n(k)$, which is obtained from (\ref{four}), is also a sum of two contributions:
$n=n_{0} + n_{1}$. This
allows a comparison with results from many body calculations in which $n_{0}$ and 
$n_{1}$ have been separately calculated. Of particular relevance are not only the 
shapes of $n_{0}$ and $n_{1}$, but also their normalizations, 
$S_{0(1)}\equiv\int n_{0(1)}d^3k = \int f_{0(1)}dy$, the so called
occupation probabilities, which, theoretically, turn out to be, for $^4He$, $S_{0} \sim 0.8$ and 
$S_{1} \sim 0.2$~\cite{panda} whereas Eq.~(\ref{seven}) yields $S_0
= 0.76$ and $S_1 = 0.24$ \cite{foot4}. A comparison between the momentum distributions obtained from 
$y$-scaling and the theoretical ones is shown in Fig. 2.
As can be seen the $n(k)$
compare very well with theoretical calculations \cite{foot3}. Thus, the physical interpretation of our
simple parametrization of $f(y)$ strongly suggests a two-component
form consisting of 
mean field and correlation contributions. In addition, we have shown that the momentum 
distributions resulting from $y$-scaling agree with theoretical ones (in agreement 
with previous results from Ref.\cite{ciofi}).

%
\begin{figure}
\vspace{.35cm}
\centerline{\rotate[r]{\epsfxsize=0.5\hsize\epsfbox{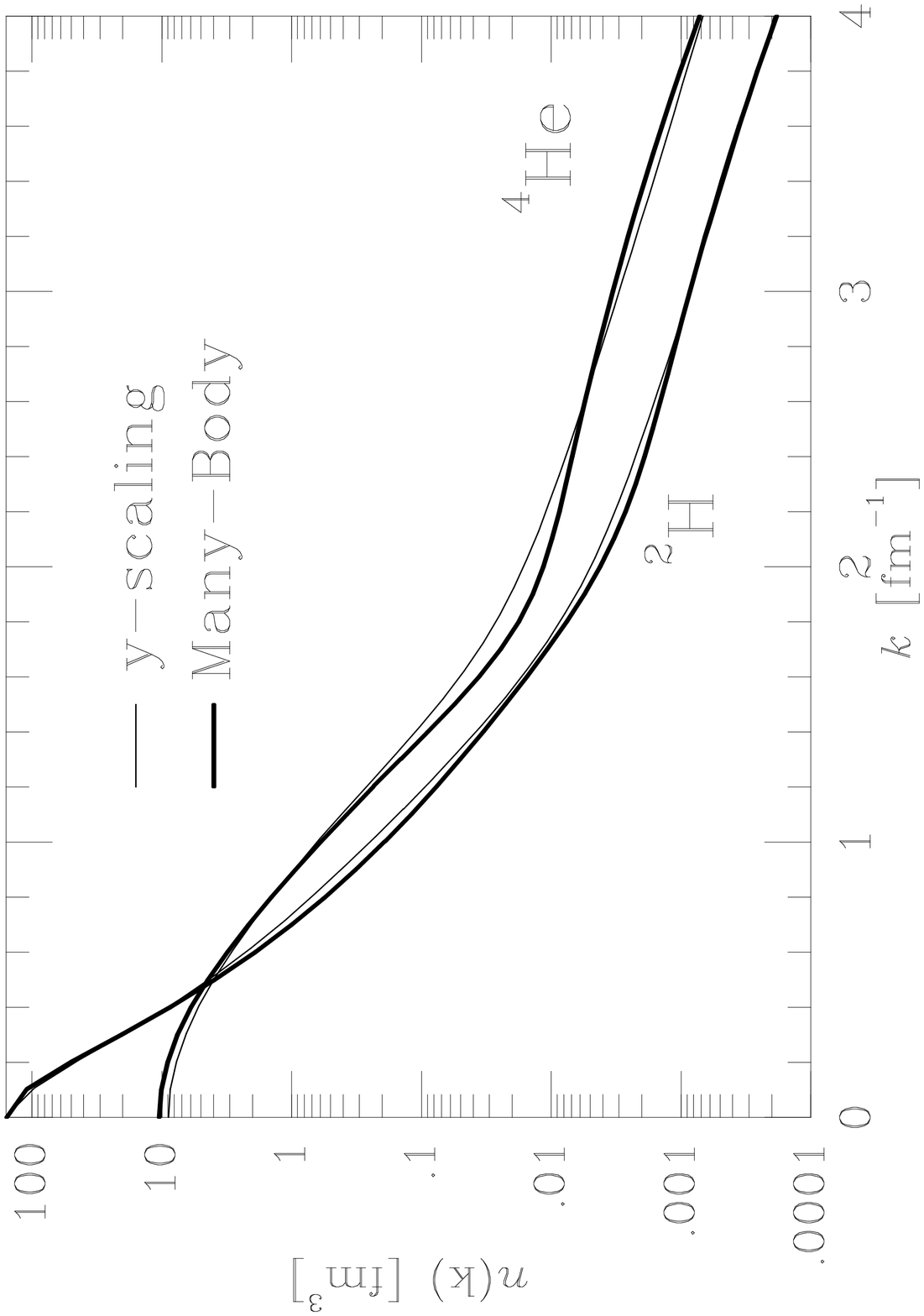}}}
\medskip

{Fig. 2. {\small{ The nucleon momentum distributions for $^2H$ and $^4He$ resulting
from Eq.~(\ref{seven}), compared with realistic momentum distributions (as parametrized in ref. \cite{frank}(b))}}}\\

\end{figure}

In making comparisons with data and/or extracting parameters for fits which can be 
compared to theoretical models a problem arises that  $f(y)$ is affected by
large  errors, particularly at large $y$. These are due to (a) experimental errors and
lack of good  inclusive data at 
large $Q^2 (\equiv \nu^2 - {\bf q}^2)$ and large $x (\equiv Q^2 /2M\nu) \ge 1$ and, (b) more to
the point here, the procedure used for applying the theoretical correction,  $B(y)$. The
first source of error could in principle be minimized  by the new generation of inclusive
data expected from CEBAF\cite{cebaf}. Here, we would like to tackle the second  problem by presenting
a new approach which would allow a determination of  $f(y)$ essentially free from
any theoretical correction contamination.

The binding correction arises from the fact that the scaling variable $y$ is 
effectively obtained from energy conservation
\begin{eqnarray}
\label{eleven}
\nu + M_{A} = [(M_{A-1} + E_{A-1}^{*})^{2} + {\bf k}^2]^{1/2} \nonumber\\ + [M^{2} + ({\bf k} + {\bf q})^{2}]^{1/2}
\end{eqnarray}
by setting $k=y$, $\frac{{\bf k}\cdot {\bf q}}{kq} = \cos \alpha = 1$, 
and, most importantly, the excitation energy, $E_{A-1}^{*}=0$; thus, $y$ represents the nucleon 
longitudinal momentum corresponding to the minimum value of the removal energy 
$(E=E_{min} + E_{A-1}^{*})$. The minimum value of the nucleon's momentum allowed 
by energy conservation in the limit $q\equiv|{\bf q}|\to\infty$, becomes 
$k_{min}(y,E)=\vert y - (E - E_{min})\vert$. It can be seen that only when $E=E_{min}$ 
does $k_{min}(y, E) = \vert y \vert$, in which case $B = 0$ and $F(y)=f(y)$. However, 
the final spectator $(A-1)$ system can be
left in all possible excited states,  including the continuum, so, in general,
$E_{A-1}^{*} \ne 0$ and $E>E_{min}$, so $B(y) \ne 0$, and $F(y) \ne f(y)$. Thus,
it is the dependence of $k_{min}$  on $E_{A-1}^{*}$ that gives rise to the binding
correction. 

%
\begin{figure}
\vspace{.35cm}
\centerline{{\epsfxsize=\hsize\epsfbox{Fig3.eps}}}
\medskip

{Fig. 3. {\small{ The experimental scaling functions $F(y)$ for $^{2}H$ and $^{4}He$  
compared with the longitudinal momentum distributions $f(y)$ given by Eq.~(\ref{seven}). In the upper part of the figure data are plotted vs. $y_1$ whereas,  in the lower part, vs. the new scaling variable, $y_2$ (see Eq.~(\ref{twelve}); for $^2H$, $y_1=y_2$). Only the data at large $q$ \cite{rock} were considered. As discussed in the text, scaling is greatly improved when the data are plotted vs. $y_2$.
}}}\\

\end{figure}

We propose  to take account of this in
the following way. We adhere to the  widespread consensus \cite{frank} that the
large $k$ and $E$ behaviours of  the nuclear wave
function are governed by configurations in which the high momentum of a
correlated nucleon (1, say) is almost entirely balanced  by another  correlated
nucleon (2, say), with the spectator $(A-2)$ system taking  only a small
fraction of $k$. Within such a picture, it can be shown \cite{frank} that  $E_{A-1}^{*}
\approx \frac{A-2}{A-1} \frac{{\bf k}^{2}}{2M}$, which is nothing but the  energy
associated with the relative motion of nucleon $2$ and system $(A-2)$. Such a  relation
represents the very physical phenomenon that in the continuum part of the  spectral
function, $E$ is a function of $k$. When this is used in Eq.~(\ref{eleven}) a new 
scaling variable, $y_{2}$, is obtained, which
incorporates the excitation energy of the final $(A-1)$ system in the  continuum\cite{foot5}: 
\begin{equation}
\label{twelve}
y_{2}={\bigg\vert} -{q\over 2} + \left [{q^2\over{4}} - \frac{4 \tilde \nu^{2} M^{2}-\rm{\tilde W}^{4}}{\rm{\tilde W}^{2}}\right ]^{1/2}{\bigg\vert}
\end{equation}
Here,
$\tilde \nu = \nu + \tilde M$, $\tilde M = 2M - E_{th}^{(2)}$, 
$E_{th}^{(2)} = \vert E_{A}\vert - \vert E_{A-2} \vert$, and $\rm{ \tilde W}^{2} 
= \tilde M^{2} + 2 \nu  \tilde M - Q^{2}$.
For the deuteron $E_{A-1}^{*}=0$, so $y_2\rightarrow y =
 \vert -q/2 + [q^{2}/4 - (4 \tilde
\nu^{2} M^{2} - \rm{\tilde W}^{4})/{ \rm{\tilde W}^{2}}]^{1/2}\vert$  
with $\tilde \nu =\nu + M_{d}$ and $M=M_d$. Thus, in general, $y_2$ can be interpreted as the
scaling  variable pertaining to a ``deuteron" with mass $\tilde M=2M-E_{th}^{(2)}$. It is 
worth stressing that for small $y_{2}( \ll (2M M_{A-1})^{1/2})$, 
$y_{2} \approx y_1$, i.e. the usual variable is recovered. Thus $y_2$ is physically
useful in both  the correlation and the single particle regions.
More importantly, since $k_{min}(q,\nu,E) \simeq \vert y_{2} \vert$, 
$B(y_2) \simeq 0$ so that  $F(y_2) \simeq f(y_2)$. Thus, plotting data in terms of
$y_2$ allows a  {\it direct} determination of $f(y_2)$. If  such a picture is
correct, one would expect from our  analysis above, the same behaviour of $f(y_2)$ at
high  values of $y_2$ for both the deuteron and complex nuclei  which is, indeed, the 
case, as exhibited in Fig.~3. This is in contrast to
what  happens with $F(y)$. 

We can summarise our conclusions as follows:
\begin{itemize}
\item[i)] The general universal features of the $y$-scaling 
function have been identified and interpreted in terms of three contributions: a
model-independent zero-range contribution, a ``universal" correlation contribution and a
mean field contribution; 

\item[ii)] The shape and evolution of the curve have been
understood both  quantitatively and qualitatively on general grounds;  

\item[iii)] By defining a proper scaling variable which  incorporates the excitation
energy of the $(A-1)$ system generated by correlations,  the longitudinal momentum
distributions can be directly obtained from the experimental data, without introducing
theoretical corrections. 
\end{itemize}

\normalsize
\vskip4cm
\huge


\begin{thebibliography}{99}

\bibitem{west} G. B. West, Phys. Rep. {\bf 18}, 263 (1975).

\bibitem{ciofi} C. Ciofi degli Atti, E. Pace and G. Salme, Phys. Rev. C{\bf 36}, 1208 (1987); {\it ibid} C{\bf 43}, 1155 (1991).

\bibitem{foot} Various corrections to this such as meson production,  off-shell effects and so on, will not be considered in
what follows since they do  not play a central role in our discussion. These will be 
discussed in a later paper; some of these corrections, such as relativistic 
effects,
can be reasonably well approximated by a suitable definition
of $y$ as in our discussion on binding corrections following Eq.~(\ref{eleven})
below.

\bibitem{day} See, for example, D. B. Day {\it et al.}, Annu. Rev. Nucl. Part. Sci. {\bf
40}, 357 (1990).

\bibitem{faralli} C. Ciofi degli Atti, D. Faralli and G. B.  West, in preparation. This will also discuss errors associated with the parameters quoted here.

\bibitem{f0} This can be expressed analytically as follows: as already implied in the above discussion, since $B\ll A/\alpha^2$ and the second term
falls off so  rapidly with $y$, the normalization integral, 
Eq.~(\ref{three}), is dominated by the small $y$ behaviour of $f(y)$, (i.e., by the
first term in Eq.~(\ref{seven})). Thus, 
$f(0) \approx (\pi^{1/2} \alpha)^{-1} = (2\pi\mu Q)^{-1/2}$.
Since the sum rule is more than $80\%$ saturated by the first term this estimate should be good to $15-20\%$ (see below). More importantly, however, 
it gives an excellent fit to the $A$-dependence of $f(0)$ and  
a simple explanation as to why
$f(0)\approx 10$ for the deuteron (and not 1, say) whereas for iron it is 
$\approx 3$ (and not 10).

\bibitem{panda} Y.Akaishi, Nucl. Phys. A{\bf 146}, 409 (1984); R. Schiavilla, V. R. Pandharipande and R. B. Wiringa, Nucl. Phys. A{\bf 449}, 219 (1986).

\bibitem{foot4} It is worth emphasizing that our values for $S_{0(1)}$ are not adjustable parameters but result from a knowledge of $f(0)$ and the behaviour of $f(y)$ at large $y$ coupled with the normalization condition, Eq.~(\ref{three}).

\bibitem{foot3} Notice that Eq.~(\ref{seven}) yields an $n(k)$ which has an unphysical 
singularity at $k=0$. This is purely an artefact of our simple parametrization of
$f(y)$ which can easily be removed by multiplying the second term in (\ref{seven}) by any
function $g(y)$ that, for small $y$, $\approx (1-by)$. A particularly convenient
choice, which introduces no new parameters and has the added virtue that it
introduces the correct $y\approx\alpha$ behaviour, is $(1 + by/(1 + y^2/\alpha^2))$. This
has almost no effect on our fits. This, and other possibilities for removing the
singularity, will be discussed in  ref.~\cite{faralli}. Note that in Fig.~2 the singularity has neen properly removed.

\bibitem{cebaf} CEBAF experiment $89-008$, B. Filippone and D. Day, spokespersons.

\bibitem{frank} (a) L. L. Frankfurt and M. I. Strikman, Phys. Rep. {\bf 160}, 235, (1988); (b) C. Ciofi degli Atti and S. Simula, Phys. Rev. C{\bf 53}, 1686, (1996).

\bibitem{foot5} From now on the usual scaling variable obtained from Eq.~(\ref{eleven}) will be denoted by $y_1$.

\bibitem{rock} S. Rock {\it et al.}, Phys. Rev. Letts. {\bf 38}, 259, (1982) and references contained therein; D. Day {\it et al.}, Phys. Rev. Letts. {\bf 59}, 427 (1987).

\end{thebibliography}
\end{document}